\documentstyle[11pt]{article}
 
\def\pa{\partial}
\def\k{\kappa}
 \def\G{\Gamma} %\mbox{\boldmath $A$}
\def\a{\alpha}
\def\b{\beta}
\def\d{\delta} 

\def\ve{\varepsilon}

\def\k{\kappa}
 
\def\m{\mu}
\def\n{\nu}

\def\s{\sigma}

\def\mn{{\mu\nu}}
\def\ab{{\alpha\beta}}
\def\be{\begin{equation}}
\def\ee{\end{equation}}

\def\bea{\begin{eqnarray}}
\def\eea{\end{eqnarray}}
\begin{document}

\begin{flushright} BRX TH-546\\BUHEP-04-13\\MIT-CTP-3533 \end{flushright}

\vspace*{.3in}

\begin{center}
{\Large\bf Cotton Blend Gravity $pp$ Waves}

{\large S.\ Deser$^1$, R.\ Jackiw$^2$, S.-Y. Pi$^3$}

{\it $^1$Department of Physics\\ Brandeis University\\
Waltham, MA 02454

$^2$Department of Physics\\ Massachusetts Institute of Technology\\
Cambridge, MA 02139

$^3$Department of Physics\\ Boston University\\ Boston, MA 02215}

\end{center}

\begin{abstract}
We study conformal gravity in $d=2+1$, where the Cotton tensor is
equated to a--necessarily traceless--matter stress tensor, for us
that of the improved scalar field.  We first solve this system
exactly in the $pp$ wave regime, then show it to be equivalent to
topologically massive gravity.
\\

Dedicated to Andrzej Staruszkiewicz, a pioneer in $d$=3 gravity,
on his 65$^{\rm th}$ birthday.
\end{abstract}

%\renewcommand{\baselinestretch}{2}
%\small\normalsize

The gravitational properties of $d = 2+1$ worlds have been studied
intensively, both in normal Einstein theory \cite{001} and in its
topologically massive extension \cite{002}.  For the latter, the
Einstein tensor $G_\mn$ is supplemented by the Cotton conformal
curvature tensor $C_\mn$ \cite{003}; the usual Weyl tensor
vanishes identically here.  Like the latter, $C_\mn$ is symmetric,
traceless and vanishes if and only if space is conformally flat.
It is also identically conserved.

Our purpose here is to examine the pure Cotton model, in which
gravity is entirely governed by  $C_\mn$, with field equations
\be%1
C_\mn = \k \; T_\mn \; ,
 \ee
 where the matter stress tensor $T_\mn$ must be traceless and
\bea%2
\sqrt{g} \; C^\mn & \equiv & \textstyle{\frac{1}{2}} \;
\ve^{\m\a\b} \, D_\a \, R^\n_\b + \textstyle{\frac{1}{2}} \;
\ve^{\n\a\b} \, D_\a \, R^\m_\b   \nonumber \\
& = & \ve^{\m\a\b} \, D_\a \, (R^\n_\b -\textstyle{\frac{1}{4}}\,
\d^\n_\b \, R) \; .
 \eea
That the two expressions for $C^\mn$ are equal follows from the
Bianchi identity.  [Our conventions are $R_\mn = +\pa_\a \:
\G^\a_\mn + \ldots$, and signature (1, -1, -1).]

We choose the simplest continuous source\footnote{In previous
studies of (1), point particle sources were considered \cite{005};
the sourceless equation, but with a dimensional Kaluza--Klein
reduction, was also solved \cite{006}.}: a conformally coupled
scalar field $\psi$ whose action is
\be%3
I = \textstyle{\frac{1}{2}} \int \, d^3x \: \sqrt{g} \: ( g^\mn \,
\pa_\m \psi \pa_\n \psi + \textstyle{\frac{1}{8}} \: R \, \psi^2 )
 \ee
with (improved) energy-momentum tensor \cite{004}
 \be%4
 T_\mn  =  \pa_\m \psi \pa_\n \psi - \textstyle{\frac{1}{2}} \,
 g_\mn g^\ab \pa_\a \psi \pa_\b \psi + \textstyle{\frac{1}{8}}\,
 \psi^2 G_\mn
  + \textstyle{\frac{1}{8}} (g_\mn D^2 - D_\m D_\n ) \psi^2
 \ee
 $T_\mn$ is covariantly conserved and traceless on the matter shell,
 \be%5
 (D^2 -\textstyle{\frac{1}{8}} \, R) \psi = 0 \; .
 \ee
We shall solve this system in the plane-fronted parallel ray
($pp$) Ansatz for the
 geometry: with $u \equiv \frac{1}{\sqrt{2}} \, (t + x), \;
 v = \frac{1}{\sqrt{2}} \, (t - x)$ ,
 \be%6
  ds^2 = F(u,y)du^2 + 2dudv -  dy^2
 \ee
so that
\be%7
g_\mn =
\begin{array}{c|ccc}
   & u & v & y \\
  \hline
u  & F & 1 & 0 \\
v  & 1 & 0 & 0 \\
y  & 0 & 0 & -1
\end{array}
\hspace{.4in} g^\mn =
\begin{array}{c|ccc}
   & u & v & y \\
  \hline
u  & 0 & 1 & 0 \\
v  & 1 & -F & 0 \\
y  & 0 & 0 & -1
\end{array}
\ee
 Note that vanishing $g_{uy}, \; g_{vy}$ and $g_{vv}$ can be
achieved by a coordinate choice, while $g_{uv}$ can be set to
unity by a conformal transformation.  Thus our Ansatz consists in
the requirement that $g_{uu}$ be $v$-independent and that $g_{yy}$
be unity.  The Ricci and Cotton tensors each possess only one
non-vanishing component
\be%8
R_{uu} = \textstyle{\frac{1}{2}} \: F^{\prime\prime} \; ,
 C_{uu} = \textstyle{\frac{1}{2}} \: F^{\prime\prime\prime}
\ee
 and $R$ vanishes, so that $G_\mn$ coincides with $R_\mn$.  (We
denote derivation with respect to $y$ by a dash; with respect to
$u$, by an over-dot.)

As is shown in Appendix A, the field equations require $\psi$ to
depend only on $u$.  We simplify our procedure by using this fact,
which implies that all energy-momentum tensor components but
$T_{uu}$ vanish. Furthermore (5) is identically satisfied.  Thus
there is only one equation to solve: $C_{uu} = \k \, T_{uu}$ or
\be%9
\frac{1}{2} \: F^{\prime\prime\prime}  = \frac{\k\psi^2}{16} \;
\left( F^{\prime\prime} + 2 \: \frac{\ddot{\s}}{\s} \right) \; ,
\hspace{.4in} \s = 1/\psi^2 \; .
 \ee
The solution is immediate,
\be%10
F(u,y) = f \, \exp [\k\psi^2 y/8] - \frac{\ddot{\s}}{\s} \: y^2 +
\a y +\b
 \ee
where $f, \; \a$ and $\b$ are three integration
constants---actually functions of $u$---arising from solving the
third-order equation (9).  Evidently the Ricci (equivalent, in
$d$=3, to the full) curvature
\be%11
 R_{uu} = \frac{1}{2} \: F^{\prime\prime} = \frac{\k^2\psi^4}{128}
 \: f \, \exp [\k\psi^2 y/8] - \frac{\ddot{\s}}{\s}
 \ee
does not depend on $(\a ,\b)$ and in Appendix B we show that a
coordinate transformation removes them.

Thus we have established that a $pp$-wave geometry is supported by
the Cotton tensor with a conformally coupled scalar field source,
which also propagates as a wave:
 \bea%12
 g_{uu} & = & f(u) \exp [\k\psi^2 y/8] - \frac{\ddot{\s}}{\s} \: y^2
 \; , \hspace{.2in} g_{uv} = 1,  \hspace{.2in} g_{yy} = -1 \nonumber
 \\[.1in]
 \psi & = & \psi (u) \; .
 \eea
Note that the scalar field is not further specified beyond
depending on retarded time, as is appropriate for a free field;
$f(u)$ is arbitrary, but its vanishing would imply that of
$C_\mn$. Also note that the exponent is proportional to $\psi^2$,
so the curvature always blows up exponentially as $\k y\rightarrow
\infty$.

The equations obeyed by a Killing vector $X_\m$ in our original
$(u,v,y)$ coordinates, where there is no $v$-dependence, require
that $X_v = {\rm const}$, $X_y = X_y(u)$ while $X_u$ obeys
\be%13
\dot{X}_u - \textstyle{\frac{1}{2}} \, F^\prime X_y -
\textstyle{\frac{1}{2}} \dot{F} X_v = 0
 \ee
\be%14
X^\prime_u + \dot{X}_y -  F^\prime X_v = 0 \; .
 \ee
For generic $f$ and $\psi$, the geometry supports only one Killing
vector: $X^\a_1 = (0,1,0)$, corresponding to a constant shift of
$v$, which clearly is an isometry of the $v$-independent metric
(12).  However, with constant $\psi$ (vanishing $\ddot{\s} /\s$)
and special form for $f$, $f = (A+Bu)^n e^{mu}$, so that
\be%15
g_{uu} = (A+Bu)^n \: \exp [mu+\k\psi^2 y/8]
 \ee
there is the additional Killing vector
\be%16
X^\a_2 = \left( - \frac{\k\psi^2}{8} \: (A+Bu) \; , \;\;
 \left( \frac{\k\psi^2_v}{8} +my\right) B \; , \;\; m (A+Bu) +
 (n+2) B \right)
 \ee
whose Lie bracket with $X^\a_1$ closes on $X^\a_1$.  Thus, since
$X^\a_1$ is a translation, $X^\a_2$ ia a dilation.  This is seen
explicitly when the following coordinate transformation is
performed (with $B\neq 0$)
 \bea%17
 U & = & A + Bu \nonumber \\[.1in]
 V & = & \frac{1}{B} \left( v+ \frac{m^2}{2a^2} \: u + \frac{m}{a}
 \, y - \frac{m^2A}{2a^2B} - \frac{2m}{a} \: \ell n \, B \right)
 \nonumber \\[.1in]
 Y & = & y + \frac{m}{a} \: u - \frac{2}{a} \: \ell n \, B \; .
 \eea
Here $a \equiv \k\psi^2 /8$.  The line element becomes
$$
ds^2 = U^n \, e^{aY} dU^2 + 2dUdV - dY^2
$$
and the dilation Killing vector reads
\be%18
X^\a = (-aU, \; aV, \; n+2) \; .
 \ee

Finally, we show that an amusing property of our CS+ scalar system
is its formal equivalence  to the CS + Einstein (=TMG) model of
\cite{002}. As is easily checked, the improved scalar's action can
be represented as the Einstein action of the rescaled metric
$g^\prime_{mn} = \psi^4 g_{mn}$,
\be%19
I[\psi ;g] = \int d^3x \sqrt{g^\prime}  R(g^\prime ) \; .
 \ee
This rescaling is purely formal: $\psi$ remains the matter field
variable. Consider now Cotton gravity; since the Cotton tensor is
conformally invariant, the gravitational field equations are
simply those of TMG,
 \be%20
C_\mn (g^\prime )= \k G_\mn  (g^\prime ) \; .
 \ee
Furthermore, the scalar's equation is already included as the
trace of (20), so variation of $\psi$ is unnecessary. Note too
that TMG, like the scalar field, has one degree of freedom, while
Cotton gravity has none. [This amusing correspondence between the
two models is no longer valid in presence of  generic matter since
traceful $T_\mn$ are forbidden here, but permitted in TMG.] The
TMG form also explains why the scalar field was rather secondary
in our $pp$ example: the only part of its stress tensor that
contributes is the improvement term $\sim R_\mn$, hence the
homogeneous nature, $F^{\prime\prime\prime} \sim
F^{\prime\prime}$, of the field equations. Since improved scalar
actions rescale to Einstein's in any $D$, the properties we have
just noted also carry over when they are coupled to the
corresponding conformal gravity models, which of course also
require traceless sources. In $D$=4, for example, we find that
adding Weyl gravity ($\int d^4x \sqrt{-g} \: C^2$) recovers Weyl
plus Einstein gravity \cite{007}.

This work was supported by the National Science Foundation under
grant PHY04-01667 and by the US Department of Energy under
cooperative research agreements DE-FC02-94-ER40818 and
DE-FG02-91-ER40676. RJ thanks C.\ Nunez for a discussion.

\newpage

\noindent{\bf Appendix A: No $v,y$ dependence of $\psi$}

\renewcommand{\theequation}{A.\arabic{equation}}
\setcounter{equation}{0}

We record the vanishing components of $T_\mn$ in (4).  As in (9),
it is convenient to work in terms of $\s = \psi^{-2}$.
\be%A.1
T_{vv} = \frac{1}{8\s^2} \, \pa^2_v \, \s  = 0
 \ee
\be%A.2
T_{vy} = \frac{1}{8\s^2} \, \pa_v \, \s^\prime = 0 \; .
 \ee
These  have the consequence that
\be%A.3
\s = g(u,y) + h (u) v \; .
 \ee
Next
\be%A.4
T_{yy} = \frac{1}{8\s^3} \: (\s\s^{\prime\prime} + \dot{\s} \pa_v
\s - \frac{1}{2}\: F \pa_v \s \pa_v \s - \frac{1}{2} \:
\s^\prime\s^\prime )\; .
 \ee
Inserting (A.3) and separating terms linear in $v$ and
$v$-independent leaves
\be%A.5
h (g^{\prime\prime} + \dot{h} ) = 0
 \ee
 \be%A.6
gg^{\prime\prime} + \dot{g} h - \textstyle{\frac{1}{2}} \: Fh^2 -
\textstyle{\frac{1}{2}} \: g^\prime g^\prime = 0 \; .
 \ee
Continuing, we examine the $uv$ component.
 \be%A.7
 T_{uv} = \frac{1}{8\s^3} \left( \s\pa_v\dot{\s} - \dot{\s}
 \pa_v \s + \frac{1}{2}\, F\pa_v \s \pa_v\s + \frac{1}{2} \,
 \s^\prime\s^\prime \right) = 0 \; .
 \ee
With (A.3) and (A.5) this requires
 \be%A.8
 g ( g^{\prime\prime} + \dot{h} ) =0 \; .
 \ee
Finally, we consider the $uu$ component equation,
\bea%A.9
T_{uu}& = & \frac{1}{8\s^3} \left( \s \ddot{\s} - F\dot{\s}\pa_v
\s
 +  \textstyle{\frac{1}{2}} \, F^2 \pa_v \s\pa_v \s +
\textstyle{\frac{1}{2}} \, F \s^\prime\s^\prime -
\textstyle{\frac{1}{2}} \dot{F} \s \pa_v \s \right. \nonumber \\
 && ~~~~~~~~~~~~~~~~~~~~~~~~ -
 \left. \textstyle{\frac{1}{2}}\, F^\prime \s\s^\prime +
\textstyle{\frac{1}{2}} \, F^{\prime\prime} \s^2 \right) =
 \textstyle{\frac{1}{2\k}} \, F^{\prime\prime\prime} \; .
\eea Upon multiplication by $\s^3$ and decomposition according to
powers of $v$, the $v^3$ term requires $h^3 F^{\prime\prime\prime}
= 0$ or $h = 0$, since we assume that the Cotton tensor is
non-vanishing.  It then follows from (A.6) and (A.8) that
$g^\prime = 0$, so $\s$ and therefore $\psi$ depend only on $u$.
Other components of $T_\mn$, as well as the matter field equation
(5), do not provide independent restrictions.

\noindent{\bf Appendix B: Removing Integration ``Constants"}

\renewcommand{\theequation}{B.\arabic{equation}}
\setcounter{equation}{0}

In the line element associated with (10)
\be%B.1
ds^2 = \left( f \exp [\k\psi^2 y/8] - \frac{\ddot{\s}}{\s} \, y^2
+ \a y +\b \right) du^2 + 2 dudv - dy^2 \; ,
 \ee
we pass to the new coordinates
 \be%B.2
 u  =  U \hspace{.5in}
 v  =  V + A(U)Y + B(U) \hspace{.5in}
 y  =  Y + C(U) \; ,
 \ee
 so that (B.1) becomes
 \bea%B.3
ds^2 & = & \left[ f \exp [\k\psi^2 C/8] \, \exp [\k\psi^2 Y/8] -
 \frac{\ddot{\s}}{\s} \: Y^2 + \left( \a - 2 \frac{\ddot{\s}}{\s}
 \: C + 2\dot{A} \right) Y \right. \nonumber \\
 & + &
 \left. \b + \a C - \frac{\ddot{\s}}{\s} \: C^2 - \dot{C}^2 +
 2\dot{B} \right] dU^2 + 2d UdV \nonumber \\
 & + & 2 \, (A-\dot{C})\, dYdU - dY^2 \; .
 \eea
 The procedure then is: with given $\frac{\ddot{\s}}{\s}$ and
 $\a$,
 solve the equation
 \be%B.4
 \ddot{C} - \frac{\ddot{\s}}{\s} \: C = - \frac{\a}{2} \; ,
 \ee
 set $A =\dot{C}$ and determine $B$ by quadrature:
 \bea%B.5
B & = & \frac{1}{2} \, \int^U du^\prime \left(\frac{\ddot{\s}}{\s}
\: C^2 + \dot{C}^2 - \a C -\b \right) \nonumber \\[.1in]
& = & \frac{1}{2} \: C\dot{C} - \frac{1}{2} \, \int^U du^\prime
\left(\frac{\a}{2} \: C + \b \right) \; .
 \eea
Upon redefining the arbitrary $f$ to absorb $\exp [\k\psi^2 C/8]$,
the line element becomes
\be%B.6
ds^2 = \left( f \exp [\k\psi^2 Y/8] - \frac{\ddot{\s}}{\s} \: Y^2
\right) dU^2 + 2dUdV - dY^2
 \ee
in agreement with (12).

\end{document}